# SynthSOD: Developing an Heterogeneous Dataset for Orchestra Music Source Separation

JAIME GARCIA-MARTINEZ [1], DAVID DIAZ-GUERRA [2] (Fellow, IEEE),
ARCHONTIS POLITIS [2] (Member, IEEE), TUOMAS VIRTANEN [2] (Fellow, IEEE), JULIO J. CARABIAS-ORTI [1],
AND PEDRO VERA-CANDEAS [1]

[1] Telecommunication Engineering Department, University of Jaen, 23071 Linares, Spain
[2] Audio Research Group, Tampere University, 33100 Tampere, Finland

CORRESPONDING AUTHOR: DAVID DIAZ-GUERRA.(e-mail: david.diaz-guerra@tuni.fi).

This work was supported by "REPERTORIUM" Project under Grant Agreement 101095065. Horizon Europe. Cluster II. Culture, Creativity and Inclusive society. Call HORIZON-CL2-2022-HERITAGE-01-02.

**ABSTRACT**   Recent advancements in music source separation have significantly progressed, particularly in isolating vocals, drums, and bass elements from mixed tracks. These developments owe much to the creation and use of large-scale, multitrack datasets dedicated to these specific components. However, the challenge of extracting similarly sounding sources from orchestra recordings has not been extensively explored, largely due to a scarcity of comprehensive and clean (i.e bleed-free) multitrack datasets. In this paper, we introduce a novel multitrack dataset called SynthSOD, developed using a set of simulation techniques to create a realistic, musically motivated, and heterogeneous training set comprising different dynamics, natural tempo changes, styles, and conditions by employing high-quality digital libraries that define virtual instrument sounds for MIDI playback (a.k.a., soundfonts). Moreover, we demonstrate the application of a widely used baseline music separation model trained on our synthesized dataset w.r.t to the well-known EnsembleSet, and evaluate its performance under both synthetic and real-world conditions.

## I. INTRODUCTION

Artificial intelligence applied to music source separation (MSS) has advanced significantly in recent years thanks to the efforts of the scientific community and the emergence of various challenges [1], [2]. In production, each instrument or voice is recorded in isolation into a separate audio track, and the mix is obtained by summing the processed tracks. This additive process is a simplification since the mix typically goes through a mastering step which includes the application of multiple non-linear transformations to the mix signal to produce the master, which is rarely a simple sum of the tracks. Nonetheless, in practice, this assumption is not an impediment to obtaining compelling results. Therefore, the target of the source separation models is to take the mix as input and estimate the inherent tracks (or stems) that compose it.

In the last decade, data-driven machine learning approaches for MSS have also been of great interest to researchers [3]. In particular, deep neural networks have been largely investigated providing a substantial improvement of the separation performance. Training supervised source separation models usually requires datasets that include clean target sources as references, which help the deep learning models learn effectively. Unfortunately, due to copyright, it is hard to obtain and share music recordings for machine learning purposes. It is even harder to obtain multi-track recordings that include isolated tracks, as these are rarely made available by artists. To overcome this limitation, the research community has nonetheless been able to create and share multi-track datasets [2], [4], [5], [6], [7]. The previously mentioned databases have contributed to significant advances in the development of separation systems in various types of music, such as pop or vocal. It is worth noting the challenges associated with the MUSDB18 database [1], [2], where algorithms are developed by the scientific community to compete under the same conditions to achieve the best separation results in a typical commercial music scenario (Voice, Drums, Bass, Accompaniment). Thanks to this effort, new models have emerged during recent years that have revolutionized the state of the art showing impressive results for this setup [8], [9], [10], [11], [12], [13].

However, in the orchestral domain, the amount of training material is very limited and no competition encourages the scientific community in the development of novel separation systems. Consequently, the several approaches that have been presented [14], [15] suffer diverse disadvantages such as limited dataset size, recording conditions, etc. It is important to recognize that acquiring individual tracks for orchestral music presents significant challenges due to the considerably larger number of instruments involved compared to other musical genres which are typically recorded at the same time and in the same room instead of instrument by instrument. Rehearsals and performances in orchestral settings are inherently synchronized and group-oriented, which makes the isolated recording of individual instrument sections a notably unnatural process.

Some data augmentation techniques have been proposed in the context of sound source separation for classical music [16], [17], [18] but none of them tries to separate every section of a full orchestra since the amount of data for this case is too low even to use data augmentation techniques on top of it. Alternatively, [19] assumes that the score is available and uses multiple synthetic renditions of the piece to train the MSS system. The work in [20] introduces a novel approach that leverages the hierarchical relationships between musical instruments to achieve more flexible and context-aware MSS, improving performance with limited training data.

Recently, a bleed-free synthetic dataset called EnsembleSet [21] was released specifically to train MSS systems to deal with ensemble music signals. Unfortunately, the amount of training material is very unbalanced (i.e. several hours for string instruments and just a few minutes of brass, woodwind, or percussion instrument sounds) and presents some limitations in terms of dynamics and velocity due to the limitations of the music information encoded in the MIDI and lilypond[1] formats.

In this paper, we present two novel contributions:
1) A large and heterogeneous dataset called SynthSOD (i.e., Synthesized Symbolic Orchestra Dataset) specifically designed for machine-learning-based orchestra music source separation systems is developed. This dataset consists of high-quality synthetic sound material covering different styles, dynamics, tempi, and techniques.

2) The proposed framework is publicly available, allowing it to be replicated or freely adapted to meet the requirements of any MSS system.

In order to test the utility of the presented dataset, it has been utilized to train a well-known state-of-the-art MSS method [22], and its performance has been compared with training using EnsembleSet and tested on real datasets of ensembles [23] and orchestra recordings [24].

## II. BACKGROUND

Training supervised source separation models requires clean target sources from datasets to guide the learning process. However, while individual performers in popular music can be recorded separately with a reference metronome or backing track, ensembles are typically recorded together in one take to maintain synchronization. This results in recordings where the audio from other instruments ("bleed") is often present, making it difficult to obtain clean stems for training. Consequently, the scarcity of clean and substantial datasets for ensembles has limited research in this area. To mitigate this drawback, several limited databases are mentioned in the literature, among which the following can be highlighted:

- TRIOS dataset [25]: This dataset consists of the bleed-free separated tracks from 5 recordings of chamber music trio pieces (4 classical and 1 jazz), with their aligned MIDI scores.
- Bach10 Dataset [26]: This dataset consists of the audio recordings of each part and the ensemble of 10 pieces of four-part J.S. Bach chorales, as well as their MIDI scores, the ground-truth alignment between the audio and the score, the ground-truth pitch values of each part, and the ground-truth notes of each piece. The audio recordings of the four parts (Soprano, Alto, Tenor, and Bass) of each piece are performed by violin, clarinet, saxophone, and bassoon, respectively. To obtain the bleeding-free signals, each musician's part was recorded in isolation while the musician listened to the recordings of others through headphones.
- MIREX multi-F0 dataset: [27]: This dataset was developed for the evaluation of multi-pitch estimation systems on the Music Information Retrieval Evaluation eXchange (MIREX) and includes the real multitrack recordings of a woodwind quintet and several synthesized extracts.
- URMP Dataset [23]: The University of Rochester Multi-Modal Music Performance (URMP) dataset comprises 44 small ensemble (2 to 5 instruments) musical pieces assembled from coordinated but separately recorded performances of individual tracks. For each piece, the musical score in MIDI format, the high-quality individual instrument audio recordings, and the videos of the assembled pieces are provided. This dataset is particularly useful for multi-modal information retrieval techniques such as music source separation, transcription, and performance analysis and also serves as ground truth for evaluating performances.

---
[1][Online]. Available: https://lilypond.org/text-input.html

- Aalto anechoic orchestra dataset [28], [29], [30]: The Aalto anechoic orchestra dataset consists of four passages of symphonic music from the Classical and Romantic periods (comprising around 10 minutes in total). This work presented a set of anechoic recordings for each of the instruments, which were then synchronized between them so that they could later be combined into a mix of the orchestra. Aligned score information and multichannel recordings of this dataset are also presented in [29] and available in [30] as part of the PHENICX project multimedia material.
- Anechoic Recording of Beethoven's Symphony No. 8 op. 93 [31]: The complete version of this dataset contains almost 20 minutes of multitrack recordings in anechoic conditions with low (but noticeable) bleeding between instruments of the same family. However, only 3 extracts of 1 minute each have been published under an open license.
- Operation Beethoven [24]: The Operation Beethoven project published the isolated recordings of the 12 sections of the first movement of Beethoven's Symphony No. 4. The recordings were conducted section by section (so they are bleed-free) by the musicians of the Hofkapelle München orchestra in a concert hall with a conventional microphone setup (so it contains a natural reverberation similar to the one we could expect from a traditional orchestra recording).
- EnsembleSet [21]: A synthetic audio dataset generated using the Spitfire BBC Symphony Orchestra library [32] and ensemble scores from RWC Classical Music Database [33] and Mutopia.[2] The data generation method introduces automated articulation mapping for different playing styles based on the input MIDI/MusicXML data. The database was rendered using 20 different mix/microphone configurations allowing the study of various recording scenarios for each performance. Overall, the dataset presents 80 tracks (6+ hours) with a range of string, wind, and brass instruments arranged as chamber ensembles.

## III. DATASET

In this work, we aim to develop a large and heterogeneous dataset of classical instrumental sounds covering different styles, dynamics, tempi, and techniques to train supervised MSS models.

The baseline score information is extracted from the Symbolic Orchestral Database[3] (SOD). SOD contains MIDI files from four different sources Musicalion (4373 files), Kunstderfuge (1466 files), Mutopia (15 files), and OpenMusicScores (14 files). Unfortunately, MIDI format is more a communication protocol rather than a format to store music information and, therefore, meaningful information such as expression marks (articulations, dynamics, and tempo marks), instrument information on the score, or note grouping cannot be directly encoded.

To overcome this limitation, we propose a musically motivated framework to automatically generate annotations for the raw MIDI files composed of several strategies that are presented in detail in this section. Finally, the annotated score synthesis is performed using a highly realistic orchestral sample library by Spitfire Audio called "BBC Symphony Orchestra" (BBCSO) [32].

The synthesized music signals[4] and the code used to extract the MIDI files from SOD, fix them to the General MIDI standard, generate the annotations, and synthesize the annotated scores are also released.[5]

### A. SELECTION OF SUBSET FROM SYMBOLIC ORCHESTRA DATASET

The MIDI files from SOD come from different origins and do not follow the General MIDI standard or any other common criteria for naming the different instruments and associating them to MIDI programs. Therefore, the first step to synthesize them was fixing the files to the General MIDI standard. To do this, we handcrafted a dictionary with the 320 most repeated instrument names in SOD and their General MIDI equivalent, which allowed us to fix more than 4,000 files.

The vast majority of the MIDI files in the SOD did not contain any explicit expressiveness notations such as articulations or dynamic markings. While some files included variations in MIDI velocity or tempo changes, these details were often inconsistent or incomplete. For instance, most MIDI files had all notes set to the same velocity and tempo values, but these values varied across files. To ensure consistency, we normalized the velocity and tempo of all notes in each fixed MIDI file to a standard velocity value of 75 and a single tempo interval of 120 beats per minute (BPM). This procedure was essential for preparing the files to be processed by our pipeline, as it ensured a consistent starting point. With this normalization, we could then apply our proposed pipeline, which introduces a controlled, consistent expressiveness to the dataset, providing the necessary diversity without compromising musical coherence.

The second step involved filtering the MIDI files to retain only those containing the orchestra instruments we were interested in and could be synthesized (e.g., excluding files with vocal tracks). Additionally, we removed duplicates, as some pieces appeared multiple times in the original SOD. Following the instruments available in the Spitfire BBC Symphony Orchestra synthesizer, we chose as our target instruments the violin, the viola, the cello, the contrabass, the flute, the piccolo, the clarinet, the oboe, the English horn, the bassoon, the French horn, the trumpet, the tuba, the harp, the timpani, and the untuned percussion (including the General MIDI instruments tinkle bell, steel drums, and percussion). In total, we found 596 unrepeated files that contained only instruments from this list and included at least two different types (i.e., monotimbral material was discarded).

---

[2][Online]. Available: https://www.mutopiaproject.org
[3][Online]. Available: https://qsdfo.github.io/LOP/database
[4][Online]. Available: https://doi.org/10.5281/zenodo.13759492
[5][Online]. Available: https://github.com/repertorium/HQ-SOD-generator

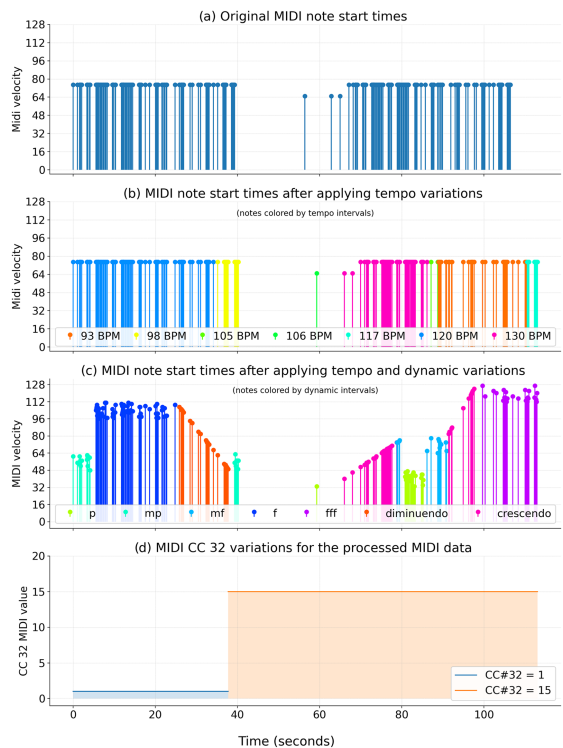

**FIGURE 1.** Representation of diverse tempo, dynamics, and articulation variations over time, illustrating the process undertaken to maximize heterogeneity in the MIDI dataset. Dynamic marks are abbreviated; see Table 1 for their full meanings.

### B. GENERATING DIVERSE AND REPRESENTATIVE CONDITIONS

While our pipeline introduces some randomness to tempo, dynamics, and articulations in the dataset's MIDI scores, this process was carefully guided by two orchestral music experts, who provided input on parameter ranges for tempo fluctuations, dynamic shifts, and articulation styles, with the articulation occurrence probabilities explicitly set by these experts, ensuring that the variations remained musically coherent and realistic. Moreover, they listened to a selection of sample outputs throughout the development process, advising on adjustments to maintain authenticity in line with real-world performance practices. By balancing randomness with expert insights, we maintain diversity without compromising the authenticity or relevance of the musical interpretations.

Fig. 1 illustrates the original data from a MIDI file and the results achieved at each stage of the proposed pipeline. Specifically, the subplots show: a) the original MIDI data, b) tempo variations, with different colors representing distinct tempo intervals. This is the first step in the pipeline, which alters the duration (x-axis) of the original MIDI file, c) dynamic changes, with colors corresponding to different dynamic intervals, applied after the tempo variations. This step modifies the MIDI velocity value compared to the input data, as reflected in the y-axis value of each note. Finally, d) represents in different colors each of the resulting articulation intervals identified by a corresponding unique MIDI CC#32 value. The color-coding visually differentiates each interval, helping to clarify how each step in the pipeline impacts the musical interpretation.

#### 1) MUSICALLY MOTIVATED RANDOM TEMPO INTERVALS
The usage of keywords, such as "largo", "lento", "andante", "allegro" or "presto", to describe tempo is a common practice in classical music. In the mentioned list, "largo" is the slowest tempo of them all, whereas "presto" is the highest tempo. However, this kind of tempo notation is not available when working with MIDI files. Instead, tempo is measured in beats per minute (BPM) which is more precise, and actually the most common way to measure tempo.

In SynthSOD, we propose that the total number of tempo intervals (i.e., segments with equal tempo values) per MIDI file to be randomly determined based on the duration of the piece, with a minimum of three tempo changes and a maximum defined by the length of the MIDI score. A tempo value is then assigned to each interval, following a normal distribution where the mean represents the average target tempo for the given MIDI file, and the standard deviation defines the tempo range. Both the mean and standard deviation are randomly computed, with default values of 120 and 30 BPM for the mean and the standard deviation, respectively.

As illustrated in Fig. 1(a) shows the original MIDI note start times, while Fig. 1(b) depicts how these note start times have been altered as a result of the tempo variations applied by our pipeline. Notes belonging to each tempo interval are color-coded differently to visually represent the tempo changes across the score.

#### 2) MUSICALLY MOTIVATED RANDOM DYNAMIC INTERVALS
In SynthSOD, different dynamic intervals have been randomly added to each processed MIDI file prior to synthesizing it. First, the number of dynamic intervals is randomly calculated in the same way as the tempo intervals. However, the number of intervals for tempo and dynamic intervals may not necessarily be the same. Although it is not possible to use the standard dynamic marks in MIDI files (forte, piano, etc.), this kind of note loudness information can be effectively designated using the MIDI velocity parameter, which consists of an integer number ranging from 0 to 127, with 0 being the "note off" MIDI message that indicates a note should stop sounding, and 127 being the loudest sound a note can be played with. Table I shows the mentioned mapping between MIDI velocity and dynamic marks.

In our framework, the MIDI velocity value for all notes within the same dynamic interval is varied by randomly selecting a value from the corresponding range. Additionally, the type of transition between adjacent dynamic intervals has also been taken into account. These transitions may be abrupt, i.e., MIDI velocity values move from one range to another at the start time of the new interval; or gradual, commonly known as diminuendos and crescendos. Gradual transitions represent a random percentage of the total amount of transitions between dynamic intervals and have been modeled as a linear function with a positive (crescendo) or negative (diminuendo) slope. The total duration of the transition is also randomly computed. These transitions can be distinguished in Fig. 1(c), where notes belonging to each dynamic interval are color-coded differently to easily visualize the introduced velocity changes.

**TABLE 1** Dynamic Marks and MIDI Velocity Mapping

| Dynamics | Mark | Meaning | Velocity |
|---|---|---|---|
| Pianississimo | ppp | very very quiet | [1,16] |
| Pianissimo | pp | very quiet | [16,32] |
| piano | p | quiet | [32,48] |
| mezzo piano | mp | moderately quiet | [48,64] |
| mezzo forte | mf | moderately loud | [64,80] |
| forte | f | loud | [80,96] |
| fortissimo | ff | very loud | [96,112] |
| fortississimo | fff | very very loud | [112,127] |

**TABLE 2** Probability Distribution and CC#32 Message Values of Articulations for String Instruments

| Articulation | CC 32 | Probability (%) | | | |
|---|---|---|---|---|---|
| | | Violin | Viola | Cello | Double Bass |
| Legato | 1 | 60.00 | 60.00 | 55.00 | 55.00 |
| Long | 2 | 0.22 | 0.37 | 0.68 | 0.86 |
| Long Con Sordino | 3 | 0.51 | 0.23 | 0.41 | 0.80 |
| Long Flautando | 4 | 0.02 | 0.44 | 0.38 | 1.21 |
| Long Harmonics | 5 | 0.42 | 0.38 | 0.74 | 1.23 |
| Long Marcato Attack | 6 | 5.00 | 5.00 | 1.05 | 0.19 |
| Long Sul Pont | 7 | 0.18 | 0.24 | 1.05 | 0.83 |
| Long Sul Tasto | 8 | 0.37 | 0.23 | 0.64 | 0.03 |
| Short Col Legno | 9 | 0.40 | 0.34 | 0.54 | 0.49 |
| Short Harmonics | 10 | 0.38 | 0.15 | 0.45 | 0.83 |
| Short Pizzicato | 11 | 0.10 | 0.10 | 5.00 | 5.00 |
| Short Pizzicato Bartok | 12 | 0.10 | 0.10 | 5.00 | 5.00 |
| Short Spiccato | 13 | 0.49 | 0.18 | 0.15 | 0.59 |
| Short Spiccato Con Sordino | 14 | 0.31 | 0.43 | 0.70 | 0.76 |
| Short Staccato | 15 | 30.00 | 30.00 | 25.00 | 25.00 |
| Tremolo | 16 | 0.41 | 0.45 | 0.88 | 0.32 |
| Tremolo Con Sordino | 17 | 0.07 | 0.33 | 0.39 | 0.12 |
| Tremolo Sul Pont | 18 | 0.48 | 0.46 | 0.66 | 0.18 |
| Trill (Major 2nd) | 19 | 0.18 | 0.21 | 0.09 | 0.9 |
| Trill (Minor 2nd) | 20 | 0.35 | 0.35 | 1.21 | 0.66 |

### 3) MUSICALLY MOTIVATED RANDOM ARTICULATIONS

Articulations specify how individual notes of a piece of music are to be performed, resulting in sound samples with distinct features. The variation of articulations within our dataset plays a critical role in ensuring its quality and applicability for training sound source separation models. By varying articulations, we aim to capture the widest possible range of sound sample conditions, thereby enhancing the model's ability to generalize across diverse musical contexts.

To fully utilize the capabilities of the BBCSO sound font library, it is essential to include every available articulation for each instrument. However, maintaining a balance that reflects real-world musical practices is crucial. Simply playing each articulation an equal number of times would create an artificially skewed dataset, where rare articulations are overrepresented and common ones underrepresented.

To ensure that the articulation selection is both musically realistic and diverse, we employ a custom weighting function for each available articulation in the BBCSO sound font library for each instrument that models the prevalence of specific articulations based on real-world musical practices, ensuring that common articulations (e.g., legato or staccato) are more likely to occur than rarer ones (e.g., trills or sul ponticello). This function was developed based on insights from expert musicians across different instrument families, modeling the prevalence of specific articulations in real-world performance practices.

In a MIDI file, each articulation corresponds to a distinct Continuous Controller (CC) 32 value. In SynthSOD, we enforce a random number of articulation intervals throughout each processed MIDI file in the same manner as it is done for dynamics and tempo variations (see Fig. 1(d)). For each interval, a CC 32 message value is assigned, corresponding to a specific articulation. The assigned value is computed based on the custom weighting function, ensuring that the choice of articulation reflects a balance between variety and real-world relevance. As an example, Table 2 presents the probability distribution for the string instruments and the associated CC 32 value. Notably, pizzicato is more common for cello and double bass compared to violin and viola whereas, rare articulations like Bartok Pizzicato and Tremolo Con Sordino have been carefully weighted to ensure they are represented without dominating the dataset. Information about the articulations of the remaining instruments in the dataset can be found in the project repository.[6]

### C. HQ SYNTHESIS

A wide range of microphones is available within the BBC Symphony Orchestra sample library [21] but, due to the size of our dataset, it is not feasible to distribute the signals of every microphone as done in EnsembleSet. Instead, we include in our dataset the Decca Tree signals, which are a reasonably good stereo mix, and the close mic signals, which are the driest signals provided by the synthesizer and could be used to simulate diverse acoustic conditions by filtering them with different room impulse responses.

We employed the Spitfire BBCSO plugin within the Reaper Digital Audio Workstation. To streamline and automate the synthesis process, a custom software tool using Reaper's Python API navigates MIDI files, assigns instrument data to tracks, sets render duration, adjusts the tempo of the project according to the MIDI data, renders audio, and finally cleans up the project. This automation significantly reduced manual

---

[6][Online]. Available: https://github.com/repertorium/HQ-SOD-generator

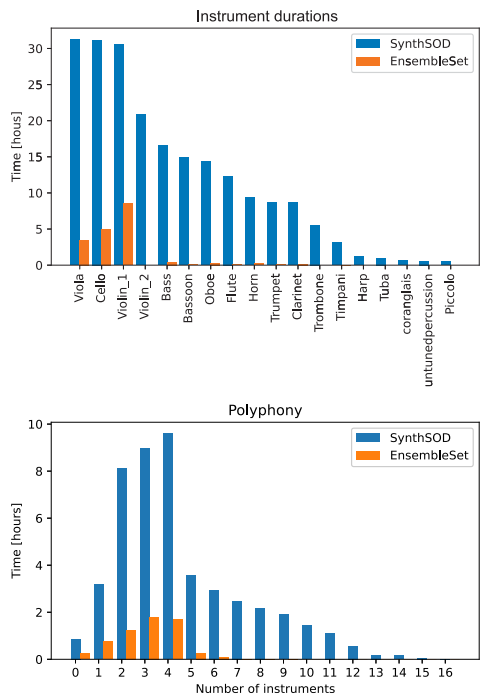

**FIGURE 2.** Activity time of every instrument (top plot) and polyphony level (bottom plot) in SynthSOD and EnsembleSet.

intervention and also enabled us to generate the dataset with precision and consistency.

The Spitfire BBCSO plugin distinguishes between long and short articulations, which results in different ways to control the dynamics. For short articulations, dynamics are managed via velocity information, while for long articulations, velocity is controlled through MIDI CC 1 (modulation wheel). To maintain consistency, we utilized a straightforward Reaper plugin[7] designed to adjust MIDI Controller CC 1 based on note velocity information.

### D. SYNTHSOD CONTENTS

After synthesizing the MIDI files, we obtained more than 47 hours of music signals, with the activity time of every instrument and polyphony level shown in Fig. 2. As we can see, the total duration of SynthSOD is more than eight times that of EnsembleSet, and the activity time of the different instruments is more balanced. Additionally, it contains a higher level of polyphony, making it more suitable for modeling full orchestras rather than just small ensembles. It is worth noting that both figures represent the times of real activity for every instrument and polyphony level, not just the sum of the duration of the pieces (i.e., a 5-minute song where the tuba is playing during 30 seconds will only contribute with 30 seconds to the tuba time and not with 5 minutes). We can see how, even if the dataset contains a large number of duos, trios, and quartets, it also contains several hours of higher polyphony situations.

Finally, we split the files into train, evaluation, and test partitions following an 70, 10, and 20% distribution and trying to keep the instrument distribution as balanced as possible using the scikit-multilearn [34] implementation of the procedures described in [35], [36].

## IV. EXPERIMENTS

### A. BASELINE MODEL FOR ORCHESTRAL MSS

As the baseline model for the dataset, we chose the open-source implementation of X-UMX [22] available in the Asteroid toolkit [37]. X-UMX is an extension of UMX [38] and was used as the baseline in the ISMIR 2021 Music Demixing (MDX) Challenge [1]. Even if more advanced models exist nowadays, we decided to use this model due to its open-source nature and baseline vocation, as well as its smaller size, which makes experimentation easier due to its lower training times. Even if the model is smaller than its most recent competitors, the GPU memory needed to train it for 15 output stems is excessive and, to make the training feasible in most consumer GPUs, we decided to train 4 independent models: one for the strings (violin, viola, cello, and bass), one for the woodwinds (flute, clarinet, oboe, and bassoon), one for the brass instruments (horn, trumpet, trombone, and tuba), and one for the percussion (harp, timpani, and untuned percussion). We joined the violin I and II stems into a single violin stem and also joined the piccolo to the flute and the cor anglais to the oboe due to their timbral similarities and their underrepresentation in the dataset.

Similar to UMX, the X-UMX model takes the magnitude spectrogram of the mixture and uses an encoder (composed of a linear layer with batch normalization and hyperbolic tangent activation), 3 BLSTM layers, and a decoder (composed of two linear layers with batch normalization and ReLU activation in the first one) to generate a spectral mask for every output stem. The main difference with UMX is the use of a bridging layer between the encoder and the recurrent layers and between the recurrent layers and the decoder. The model was trained with the combination loss function described in [22] following all the hyperparameters of the original Asteroid implementation.

The code to train and evaluate the baseline models and the pretrained models can be found in our GitHub repository.[8]

### B. RESULTS

Apart from training the baseline model with the Decca Tree signals of SynthSOD, we also trained it on EnsembleSet to prove the advantages of having a larger amount of data with more instruments and a higher level of polyphony. We evaluated the models on the test partition of SynthSOD and on the real-world URMP and the Operation Beethoven recordings.

We evaluated the baseline model by computing the signal-to-distortion ratio (SDR) of the different output stems. We first computed the SDR on one-second frames and then computed the median value for every music piece (ignoring the frames where the reference signal was silent) and finally computed

---

[7][Online]. Available: https://forum.cockos.com/showthread.php?t=21320

[8][Online]. Available: https://github.com/repertorium/SynthSOD-Baseline

TABLE 3 Signal to Distortion Ratios [Db] for the Baseline Model Trained With SynthSOD and EnsembleSet Evaluated in Different Datasets

| Evaluation on: | SynthSOD | | | EnsembleSet | | Operation Beethoven | | | URMP | | |
|---|---|---|---|---|---|---|---|---|---|---|---|
| | | Training on | | | Training on | | Training on | | | Training on | |
| Instrument | Original | SynthSOD | EnsembleSet | Original | SynthSOD | Original | SynthSOD | EnsembleSet | Original | SynthSOD | EnsembleSet |
| Violin | -12.35 | 3.66 | 0.46 | -3.37 | 5.81 | -5.00 | 1.62 | 1.65 | -2.79 | 1.09 | 1.06 |
| Viola | -10.05 | 2.92 | 0.80 | -12.77 | 0.64 | -16.21 | 0.00 | -1.71 | -4.56 | 1.23 | 0.48 |
| Cello | -4.51 | 4.90 | 2.20 | 1.00 | 8.94 | -11.00 | -0.06 | -3.00 | -4.24 | 4.51 | 2.39 |
| Bass | -7.02 | 6.24 | 3.16 | -9.10 | 4.61 | -11.53 | 0.57 | 0.50 | -6.57 | 0.21 | 0.04 |
| Flute | -17.26 | 0.48 | 0.00 | -5.75 | 3.27 | -19.85 | 0.00 | -2.88 | -2.36 | 0.95 | 0.17 |
| Clarinet | -15.36 | 0.05 | -0.07 | 0.68 | 3.04 | -15.49 | 0.00 | -0.38 | -4.79 | 0.04 | 0.22 |
| Oboe | -14.15 | 2.06 | 0.07 | -6.88 | 2.50 | -19.53 | 0.00 | -0.48 | -7.06 | 0.03 | 0.27 |
| Bassoon | -15.21 | 0.35 | 0.00 | -5.22 | 2.30 | -15.86 | 0.00 | 0.00 | -5.06 | 0.03 | 0.00 |
| Horn | -20.00 | 0.12 | -3.25 | -1.17 | 0.86 | -17.25 | -0.50 | 2.81 | -6.50 | 0.24 | 0.77 |
| Trumpet | -6.52 | 3.94 | 0.23 | 2.26 | 6.31 | -24.81 | -0.15 | -11.38 | -2.63 | 0.36 | 0.04 |
| Trombone | -4.23 | 6.15 | 0.00 | | | | | | -4.23 | 0.01 | 0.00 |
| Tuba | -0.75 | 2.56 | 0.00 | | | | | | -5.65 | 0.00 | 0.00 |
| Harp | -15.36 | 0.00 | -0.03 | | | | | | | | |
| Timpani | -24.88 | 0.29 | -7.81 | -10.02 | 9.11 | -19.40 | 0.01 | -6.40 | | | |
| Unt. perc. | -24.76 | 0.00 | -2.52 | | | | | | | | |
| MEAN | -12.83 | 2.25 | -0.45 | -4.58 | 4.23 | -15.99 | 0.14 | -1.93 | -4.70 | 0.72 | 0.45 |

The first column of every evaluation dataset indicates the SDR of the original mixtures for every instrument. Note that EnsembleSet does not include train and test partitions, so the model trained on Ensembleset cannot be evaluated on it.

TABLE 4 Signal to Distortion Ratios [Db] for the Baseline Model Trained With SynthSOD and EnsembleSet Evaluated in the Ensembles (Up to 5 Instruments) and Orchestras (More Than 5 Instruments) of the Test Partition of SynthSOD

| Evaluation on: | Ensembles in SynthSOD | | | Orchestras in SynthSOD | | |
|---|---|---|---|---|---|---|
| | | Baseline trained on | | | Baseline trained on | |
| Instrument | Original | SynthSOD | EnsembleSet | Original | SynthSOD | EnsembleSet |
| Violin | -6.77 | 5.48 | 2.74 | -16.77 | 1.02 | -2.25 |
| Viola | -7.22 | 5.72 | 2.10 | -11.69 | 1.47 | 0.13 |
| Cello | 0.07 | 10.15 | 5.57 | -7.49 | 3.07 | 1.04 |
| Bass | -7.23 | 6.81 | 3.42 | -7.02 | 6.20 | 3.16 |
| Flute | -14.24 | 1.92 | 0.00 | -17.30 | 0.47 | -0.01 |
| Clarinet | -11.38 | 0.15 | 0.00 | -15.60 | 0.05 | -0.14 |
| Oboe | -8.26 | 7.74 | 1.53 | -15.43 | 1.41 | 0.05 |
| Bassoon | -8.75 | 3.90 | 0.04 | -15.88 | 0.28 | 0.00 |
| Horn | -12.90 | 0.48 | -7.01 | -20.43 | 0.05 | -2.78 |
| Trumpet | -0.98 | 9.57 | 0.44 | -13.60 | 0.75 | 0.01 |
| Trombone | 0.33 | 8.60 | 0.01 | -14.89 | 0.00 | -0.07 |
| Tuba | 4.06 | 4.26 | 0.00 | -4.27 | 1.52 | 0.00 |
| Harp | -11.56 | 1.49 | 0.17 | -16.51 | 0.00 | -0.19 |
| Timpani | -24.38 | 1.33 | -3.91 | -24.92 | 0.25 | -8.08 |
| Unt. perc. | | | | -24.76 | 0.00 | -2.52 |
| MEAN | -7.80 | 4.83 | 0.37 | -15.04 | 1.10 | -0.78 |

The first column of every evaluation dataset indicates the SDR of the original mixtures for every instrument.

the median over the music pieces (ignoring those where the reference signal was silent). This evaluation was performed using the fourth version of the museval library [39], which was used in the 2018 Signal Separation Evaluation Campaign (SiSEC 2018) [40] and allows for some level of linear distortion in the estimates.

As we can see in Table 3, sound source separation is an extremely challenging task in orchestral music, with most of the instruments having an SDR lower than −10 dB in the original recordings. Despite this, the baseline model trained on SynthSOD is able to achieve SDRs above 2 dB for all the string instruments, all the brass instruments (except the horn), and the oboe. On the other hand, the same model trained on EnsembleSet is only able to obtain these results for the cello and the bass (and, even in those, they are clearly below the results of the model trained with SynthSOD).

When evaluated on EnsembleSet, the model trained on SynthSOD achieved quite good results. Even the woodwinds, which did not perform well on SynthSOD, showed clearly positive SDRs. This suggests that the model did not overfit to the automatic annotations generated for SynthSOD and generalizes well to real, manually created annotations by professionals. Since EnsembleSet does not include an official train and test partition, we used the whole dataset for training and, therefore, the baseline model trained with Ensembleset cannot be evaluated on it.

SynthSOD contains both ensembles and orchestras. Consequently, we divided the test partition based on the number of instruments and evaluated the SDRs of every case separately. As we can see in Table 4, for the ensemble pieces (up to 5 instruments), the model trained on SynthSOD obtains relatively good SDRs for all the instruments except for the clarinet, the horn, and the percussion, while the orchestra pieces (more than 5 instruments) are definitely more challenging. Surprisingly, the model trained on EnsembleSet does not generalize really well to the ensembles in SynthSOD, probably because of the higher diversity in terms of instruments and because the model had overfitted due to the small size of EnsembleSet.

**TABLE 5** Average Signal to Distortion Ratios [Db] of the String Instruments for the Baseline Model Trained and With the EnsembleSet Files Without Any Tempo, Velocity, and Articulation Annotations (Plain), With Our Proposed Random Annotations (Proposed), and With the Original Human-Made Annotations (Original)

| Training | Evaluation | | | |
|---|---|---|---|---|
| | Plain | Proposed | Original | URMP |
| Plain | 7.52 | 5.33 | 3.02 | 1.55 |
| Proposed | 7.13 | 6.96 | 5.22 | 1.92 |
| Original | 3.06 | 3.93 | 8.92 | 1.40 |

Finally, we evaluated the models on real recordings to determine whether they could generalize to actual instruments after being trained on synthesized signals. As shown in Table 3, none of the models were able to achieved strong separation results even in URMP, which is a less challenging dataset containing only ensembles of up to 5 instruments. It is evident that specific fine-tuning or domain-adaptation strategies [21], [41] are needed for models trained on synthesized signals to generalize effectively to real recordings.

### C. MUSICALLY MOTIVATED RANDOM ANNOTATIONS

In order to analyze the effectiveness of the process to generate the musically motivated annotations presented in Section III-B, we resynthesized the MIDI files of the EnsembleSet dataset, both without annotations (here denoted as Plain) and with the proposed musically motivated random annotations. After that, we trained the baseline model using 70% of the synthesized files from each case (including the original Ensembleset files with real human annotations) and evaluated the performance of each trained model in the remaining 30% of the files from each case, as well as on the URMP dataset.

Table 5 shows the average SDR of the string instruments (the ones best represented in the dataset) for every training and test combination. We can see how the model trained with our proposed musically motivated random annotations generalizes better to the original annotations than the model trained without any kind of annotations. Interestingly, the model trained with the original dataset does not generalize well to the other cases, probably because it is less diverse. As we could expect from the results of Section IV-B, none of the models generalize well to the real recordings of URMP, although the model trained with the proposed technique obtains slightly better results than the one trained with the other alternatives.

### V. CONCLUSION

We have presented a new dataset for MSS in the orchestral domain and the first results on the topic using neural networks. This new research problem presents many challenges compared with other MSS domains, such as a higher number of instruments and the sparsity of some of them, a higher dynamic range and spectral overlap, or the scarcity of real recordings with isolated stems for every instrument.

The baseline model obtained good separation results for most of the instruments when evaluated with synthetic data and we could expect them to be improved when using more advanced models, but the main challenge is its generalization to work with real recordings. We believe that the main reason for this is that the Spitfire BBC Symphony Orchestra library uses (as most synthesizers) a limited amount of recordings of every note and repeats them on a round-robin basis. This provides reasonable results in terms of audio quality but limits the actual diversity of the dataset even when its total duration is increased.

With the provided code, it would be possible to resynthesize the MIDI files with different synthesizers and combine them to obtain a more diverse training material. Specific fine-tuning and domain-adaptation techniques could be explored in the future, such as the ones presented in [21], [41] for small ensembles. We expect this dataset to enable the music information retrieval and signal processing communities to start exploring this new problem and allow obtaining better and more robust results in the following years.

The presented dataset is much larger and more diverse than any other dataset for MSS of classical instruments and we have presented a new procedure to automatically generate annotations for the raw MIDI files. The presented experiments prove that models trained with these automatically-generated annotations generalize well to signals synthesized using hand-made annotations done by experts.

Apart from the synthesized music signals, we are also publishing the code used to extract the MIDI files from SOD, to generate the annotations, and to synthesize them. Therefore, further researchers can generate their own versions of the dataset by, for example, choosing other instruments of interest or using different synthesizers.

### ACKNOWLEDGMENT
Authors thank John Anderson and Thomas Vingtrinier for their advice on the design of the process to generate musically motivated annotations for the MIDI files. The authors wish to acknowledge CSC—IT Center for Science, Finland, for computational resources.